# Optical Aharonov-Bohm effect in stacked type-II quantum dots

Igor L Kuskovsky,[1][*] W. MacDonald,[1] A. O. Govorov,[2] L. Muroukh,[1] X. Wei,[3] M. C. Tamargo,[4] M. Tadic,[5,6]

F. M. Peeters[5]

[1]*Department of Physics, Queens College of CUNY, Flushing NY 11367*

[2]*Department of Physics and Astronomy, Ohio University, Athens OH 45701*

[3]*NHMFL, Tallahassee, FL*

[4]*Department of Chemistry, The City College of CUNY, New York, NY 10031*

[5]*Department of Physics, University of Antwerp, Groenenborgerlaan 171, B-2020 Antwerp, Belgium*

[6]*Faculty of Electrical Engineering, University of Belgrade, PO Box 3554, 11120 Belgrade*

(12 June 2006)

Excitons in vertically stacked type-II quantum dots experience the topological magnetic phase and demonstrate the Aharonov-Bohm oscillations in the emission intensity. Photoluminescence of vertically stacked ZnTe/ZnSe quantum dots is measured in magnetic fields up to 31 T. The Aharonov-Bohm oscillations are found in the magnetic-field dependence of emission intensity. The positions of the peaks of the emission intensity are in a good agreement with numerical simulations of excitons in stacked quantum dots.

---

[*] Corresponding author. Email address: Igor.Kuskovsky@qc.cuny.edu

The wave function of a quantum particle moving in a magnetic field will experience a phase shift proportional to the magnetic flux, $\Phi$, and the experimentally observable effects will be periodic with the period equal to the flux quantum, $\Phi_0 = h/e$. This is the Aharonov-Bohm (AB) effect [1]. The AB effect is typically observed as interference of a quantum charged particle moving along a closed trajectory in a magnetic field. Experimentally, many manifestations of the AB effect for charged particles have been observed. Optically, a signature of the AB effect has been reported for the etched quantum rings [2] where an appearance of the AB effect is natural due to a non-zero total charge of a trion. Overall, the growth and magneto-optical characterization of epitaxial quantum rings are very active fields of research [3] and the AB effect is one of the central topics in these studies. Also, carbon nanotubes can demonstrate the AB features [4] optically, however small radii of nanotubes do not allow for observation the first AB oscillation [4].

One of the most interesting problems in crystalline quantum rings is whether an overall neutral particle can exhibit the experimentally observable AB effect [5]. Theoretical studies (see e.g. Refs. 6 - 9) predict that such an effect indeed should be observed in the optical emission energy of neutral excitons in both quantum rings and type-II quantum dots. Moreover, in addition to the changes in the excitonic energy, it is expected that the AB phase will reveal itself through changes in the intensity of the photoluminescence (PL), which can be quenched in the magnetic field that corresponds to the transition of the exciton angular momentum to a non-zero value [6 - 9]. The AB oscillations of the energy of such excitonic emission on the magnetic field have been reported for type-II InP/GaAs self-assembled QDs [10]; however, the PL intensity dependence on the magnetic field has not yet been reported and not well understood.

In this Letter we present the results of experimental and theoretical studies on magneto-excitons in type-II quantum dots (QDs) formed in the Zn-Se-Te multilayer system that contains ZnTe/ZnSe type-II quantum dots separated by relatively thin, nominally undoped, ZnSe barriers [11]. The typical width of the barriers is 3nm; the tellurium composition within the barriers is less than 2%, whereas the Te fraction in the QDs is typically 50 –70% [12]. In ZnTe/ZnSe QDs the hole is strongly confined within the ZnTe-rich dot due to the larger valence band offset (0.8 – 1.0 eV) [13, 14], whereas the electron is located within the ZnSe-rich barriers.

We report here, for the first time experimentally, that the integrated PL intensity exhibits an oscillatory behavior as a function of the magnetic field, as was predicted for a neutral exciton in type-II QDs [6 - 9]. Theoretically, we explain this as a motion of an electron around a entire stack of QDs, one of which is occupied by a hole. This suggests that a magnetic field can be used to control optical emission of type-II QDs as well as other type-II nanostructures, which can be useful for quantum information-related applications.

This system has several important differences from the one studied by Rebiero *et al.* [10]. First, the stacked cylindrical geometry nicely defines the ring-like trajectory for an electron, ensuring that the electron's wave-function is "pushed" to the side of the dot, due to electron-electron interaction, independent of the stress in the system (see Fig. 1(a)). Indeed, the electron density calculations for a single dot show that the electron will be located either above or below the dot, in the absence of strain, and, therefore, no AB signature is expected. Second, a free particle here is an electron, which has a smaller effective mass.

The exciton energies in stacks of ZnTeSe/ZnSe quantum dots are computed within the single band effective-mass model for both the conduction and the valence band. A similar model was recently adopted to compute the electronic structure of type-II InP/InGaP quantum dots and quantum-dot molecules [8, 15], which we extend here to model a superlattice of disk-shaped quantum dots. The strain is taken to be constant inside the dot and zero in the spacer, which is a reasonable approximation for flat shaped quantum dots. The relevant strain tensor components are extracted from the continuum mechanical model and are used in all dots. Because the spatial confinement and strain lift the energies of the heavy-hole states with respect to the light holes, only the heavy-hole exciton states are populated at $T = 4.2$ K. The diagonal approximation to the multi-band Hamiltonian is applied for the heavy hole and, therefore, different effective masses are used for the radial and vertical directions 16]. Since the size and composition of the dots in the stack are not fully controlled, a certain discrepancy between the dot radii is assumed, but all dots are assumed to have the same height and spacer thickness.

Our model is shown schematically in Fig. 1(a). From our numerical simulations we see that tunneling of the hole is very weak and therefore the exciton should be trapped in one QD. In addition, it is likely that the radii of the QDs are different. To compute an exciton trapped in an infinite stack, we model a stack as a superlattice with a relatively long period. Each period in the stack consists of a single large dot of radius $R$ and seven smaller dots of radius $R_1 = R - \Delta$ (for comparison with experiment, we took $\Delta = 1$ nm and $R = 11$ nm). The exciton is trapped in the larger dot.

In order to find the single-particle states, the finite element method is employed to solve the Schrödinger equation. The exciton states are computed as follows. The Hartree iteration is used first to deliver a suitable basis for a subsequent exact diagonalization procedure. The basis in the exact diagonalization calculation is formed as the product of five electron wavefunctions of orbital momenta $l_e$ and six hole wavefunctions of orbital momenta $l_h$, such that $L = l_e + l_h$ is the orbital momentum of the exciton, which is a conserved quantity and is fixed. In addition to the degeneracy with respect to $L$, the exciton states are arranged in quartets of spin degenerate states in the absence of the magnetic field. Among the four states in the quartet only those with equal spins of the electron and the hole are optically active.

The dependence of the band gap in $ZnTe_xSe_{1-x}$ on x is taken from Ref. 17 (we note that other models do exist; however, actual values would not affect our final results); furthermore, we assumed that for the dots $x = 0.6$. The experimental transition energy $E_{exc} \approx 2.46$ eV is reproduced by selecting the conduction band offset of the strained structure $\Delta E_C = 333$ meV, while the valence band offset is $\Delta E_V = 770$ meV in good agreement with previously reported unstrained values [13, 14]. The first angular momentum transition is found at $B_1 \approx 1.8$ T, when the radius of the large dot is taken to be $R = 11$ nm (Fig. 1(b)). The thickness of the dot was chosen to be equal to the thickness of the dot layer $h = 0.7$ nm, and for the spacer thickness we took the measured value [12] $s = 2.42$nm. The parameters of the band structure and the permittivity in ZnTe and ZnSe are taken from Ref. 14, while the elastic constants and deformation potentials are taken from Ref. 18.

Fig. 1(b) shows the variation of the lowest spin-down electron states of different orbital momenta with magnetic field. The hole is confined in a QD due to the built-in nanostructure potential while the electron is confined only due to the Coulomb interaction to the hole. The electron is distributed over a cylindrical surface, while the hole is strongly confined inside a QD (Fig. 1(d)). This spatial distribution of the particles determines the characters of *B*-dispersion of the particle energies in the magnetic field. Figure 1(c) shows that, opposite to the electronic states, no angular momentum transitions are found for the heavy hole ground state. Both spin-up and spin-down hole states are shown (Fig. 1 (c)). The heavy hole ground state has $l_h = 0$ and the wavefunction is peaked in the center of the large dot, and its probability distribution is almost unaffected by the magnetic field. The smaller dots in the stack are not populated. Even though the chosen $\Delta = 1$ nm is only 10% of the radius of the large dot, the hole is localized in the large dot in each period. In order to illustrate the difference between the electron and the hole localization in the structure, we plot in Fig. 1(d) the averages of the ground state probability densities of the electron and hole ground states over the *z*-direction. These averages are scaled to span the same range.

The lowest energy levels of spin-down exciton states are shown in Fig. 2(a) as a function of the magnetic field. The spin-up states have a similar variation, except that they have higher energies than the spin-down states. The states displayed in Fig. 2(a) exhibit several angular momentum transitions in the experimental range (see below) from 0 to 31 T. In the axially symmetric structure, only the $L = l_e + l_h = 0$ states are optically active which would result in zero PL intensity beyond the first electron transition at $B_1$ = 1.8 T; note that the hole does not exhibit transitions. However, any deviation from perfect axial symmetry may turn the $L > 0$ states optically active [19]. For example, such deviations can appear for elongated QDs or in the presence of structural defects. Note that self-assembled QDs are typically elongated due to the specific symmetry of the sample surface.

The computed binding energy of the exciton state at zero magnetic field equals 7.3 meV, which is in excellent agreement with the measured 6meV to 8meV [11, 20], and as shown in Fig. 2(b), the binding energies of all states increase with magnetic field. The calculated dependence of the exciton energy levels shown in Fig. 2(a) is typical for quantum rings and type-II quantum dots [8]. We note here that two subsequent angular momentum transitions at low magnetic field are separated by approximately 3.6 T, while this difference increases to 6T for $B \approx 27$ T. This is a clear signature that the effective exciton radius decreases with magnetic field. At low magnetic fields,

$$R_{eff} = \sqrt{\frac{\Phi_0}{2\pi B_1}} = 19.1 \text{nm}, \qquad (1)$$

where $B_1 = 1.8$ T is the magnetic field of the first angular momentum transition. At $B = 27$ T, where the sixth angular momentum transition takes place, we find $R_{eff} = 16.4$ nm.

Experimentally, magneto-photoluminescence (magneto-PL) within the Faraday configuration up to 31 T on a series of ZnTe/ZnSe multilayer QD samples has been performed at the National High Magnetic Field Laboratory. For the excitation, the 351 nm line of an $Ag^+$ laser as well as the 325 nm line of a HeCd laser was used. The PL was detected using a 1 m monochromator coupled with a liquid nitrogen cooled CCD camera. All experiments were performed at $T = 4.2$ K. A typical magneto-PL from one of the samples (Sample 1), for several values of the magnetic field and two excitation intensities is shown in Figs. 3(a) and 3(b). The spectra are essentially the same as those reported, previously, for these type-II excitons [11], including the expected excitation intensity dependent spectral position (see Ref. 11 and references therein). We note (not shown here) that the PL intensity is a linear function of the excitation intensity for several orders of magnitude, suggesting that most of the QDs are not occupied as expected within our model. Next, in Fig. 4 we plot the integrated PL intensity as a function of the magnetic field for the spectra (Sample 1) obtained under the lowest excitation, where the PL from quantum dots dominates [11]. The overall integrated intensity decreases due to magnetic-field-induced carrier localization as has been previously observed for type-II multiple quantum well structures [21]. The major feature of the spectrum is the presence of a strong peak in the intensity at $B = 1.79 \pm 0.03$ T [22]. The observed behavior is reversible, i.e. photo-degradation or other photo-chemical processes can be excluded. We thus argue that this peak is due to the AB phase as has been predicted earlier and is in excellent agreement with the results of numerical calculations for this system as described above. Next, we consider the effects of the Te fraction on the size of the islands. Gu *et al*. [20] have shown that the Zn-Se-Te multilayer samples grown with higher Te flux have larger quantum dots. Our data are consistent with such a conclusion. Figure 5 shows the data on three samples grown with increasing Te fraction. Clearly, the magnetic field value required for the orbital momentum transition decreases with increasing Te concentration, indicating the presence of laterally larger quantum dots.

Finally, we comment on the fact that PL does not vanish with increasing magnetic field and on the second feature observed at $B \approx 3.1$ T in Sample 3. For ideal QDs with cylindrical symmetry, the theory predicts that PL transitions become forbidden at $B > B_{crit}$, where $B_{crit} = B_1$ is the magnetic field of the first ground-state transition for the electron ($L = 0 \rightarrow L = -1$). Real QDs always have defects and are usually elongated. In the case of non-cylindrical dots, the selection rules are relaxed and the dipole emission is allowed for an arbitrary $L$ (see e.g. Ref. 19), so more features in the dependence of the PL intensity on the magnetic field become possible. This case was theoretically calculated using a strongly simplified 1D-ring model in Ref. 19. In particular, it was shown in Ref. 19 that PL intensity does oscillate with magnetic field, especially at non-zero temperature, and the maxima in the AB oscillations correspond to the ground state transitions in the exciton energy. The latter comes from the fact that higher energy exciton states become involved in the PL process. Our experiments demonstrate such oscillations; moreover, the flux values of the oscillations are in good agreement with the numerical data shown in Figs. 1(b) and 2(a). We assume that peaks in the magneto-PL intensity correspond to magnetic fields at which different exciton ground states cross each other in Fig. 2(a).

In conclusion, our model of the exciton states in ZnTeSe/ZnSe quantum dots indicates that the wavefunction of the electron is localized around the radial periphery due to the joint action of Coulomb interaction with the hole and the repulsive QD potential. Because of the smallness of the spacer layer between the quantum dots, the electron is unable to penetrate in the region between the dots resulting into a favorable situation for the occurrence of the Aharonov-Bohm oscillations in the electron energy levels. Our magneto-PL data taken from several samples clearly show a non-monotonous behavior of the exciton-emission intensity as a function of the magnetic field, which is a manifestation of the AB effect, and our numerical calculations support such an interpretation.

This work was supported in part by the EU-NoI: SANDiE, the Belgian Science Policy, and the Ministry of Science and Environmental Protection of the Republic of Serbia. A portion of this work was performed at the NHMFL, which is supported by NSF Cooperation Agreement No. DMR-0084173 and by the State of Florida.

**REFERENCES**

1. Y. Aharonov and D. Bohm, Phys. Rev. **115**, 486 (1959).

2. M. Bayer, M. Korkusinski, P. Hawrylak, T. Gutbrod, M. Michel, and A. Forchel, Phys. Rev. Lett. **90**, 186801 (2003).

3. P. Offermans, P. M. Koenraad, J. H. Wolter, D. Granados and J. M. Garcia, V. M. Fomin, V. N. Gladilin, and J. T. Devreese, Appl. Phys. Lett. **87**, 131902 (2005); T. Kuroda, T. Mano, T. Ochiai, S. Sanguinetti, K. Sakoda, G. Kido, and N. Koguchi, Phys. Rev. B **72**, 205301 (2005); B. Alen, J. Martinez-Pastor, D. Granados, and J. M. Garcia, Phys. Rev. B **72**, 155331 (2005).

4. S. Zaric, G. N. Ostojic, J. Kono, J. Shaver, V. C. Moore, M. S. Strano, R. H. Hauge, R. E. Smalley, and X. Wei, Science **304**, 1129 (2004).

5. A. V. Chaplik, JETP Lett. 62, 900 (1995); B. C. Lee, O. Voskoboynikov, and C. P. Lee, Physica E **24**, 87 (2004).

6. A. V. Kalameitsev, A. O. Govorov, and V. Kovalev, JETP Lett. **68**, 669 (1998).

7. A. O. Govorov, S. E. Ulloa, K. Karrai, and R. J. Warburton, Phys. Rev. B **66**, 081309(R) (2002).

8. K. L. Janssens, B. Partoens, and F. M. Peeters, Phys. Rev. B **64**, 155324 (2001).

9. I. Climente, J. Planelles, and W. Jaskolski, Phys. Rev. B **68**, 075307 (2003).

10. E. Ribeiro, A. O. Govorov, W. Carvalho, Jr., and G. Medeiros-Ribeiro, Phys. Rev. Lett. **92**, 1264021 (2004).

11. Y. Gu, Igor L. Kuskovsky, M. van der Voort, G. F. Neumark, X. Zhou, and M. C. Tamargo, Phys. Rev. B **71**, 045340 (2005).

12. Private Communications. The results are obtained from high resolution symmetrical and asymmetrical X-ray diffraction and will be published elsewhere. The data on a similar system can be found in Gong et al., J. Appl. Phys. **99**, 064913 (2006).

13. F. Malonga, D. Bertho, C. Jouanin, and J.-M. Jancu, Phys. Rev. B **52**, 5124 (1995).

14. S.H. Wei and A. Zunger, Phys. Rev. B **53**, 10457 (1996).

15. M. Tadić, F. M. Peeters, and K. L. Janssens, Phys. Rev. B **65**, 165333 (2002).

16. S. Lee, F. Michl, U. Rossler, M. Dobrowolska, and J. K. Furdyna, Phys. Rev. B **57**, 9695 (1998).

17. M. J. S. P. Brasil, R. E. Nahory, F. S. Turco-Sandroff, H. L. Gilchrist, and R. J. Martin, Appl. Phys. Lett. **58**, 2509 (1991).

18. J. Sörgel and U. Scherz, Eur. Phys. J. B **5**, 45 (1998).

19. L. G. G. V. Dias da Silva, S. E. Ulloa, and A. O. Govorov, Phys. Rev. B **70**, 155318 (2004).

20. Y. Gu, Igor L. Kuskovsky, M. van der Voort, G. F. Neumark, X. Zhou, M. Munoz, and M. C. Tamargo, Phys. Stat. Sol. (b) **241**, 550 (2004).

21. Trüby, M. Potemski, and R. Planel, Solid-State Electron. **40**, 139 (1996); M. Haetty, M. Salib, A. Petrou, T. Schmiedel, M. Dutta, J. Pamulapati, P. G. Newman, and K. K. Bajaj, Phys. Rev. B **56**, 12 364 (1997).

22. No such peak is observed at high excitation intensity (Fig. 4, inset (b)), since it is masked by the emission due to isoelectronic bound excitons [11], which always exist in Zn-Se-Te systems.

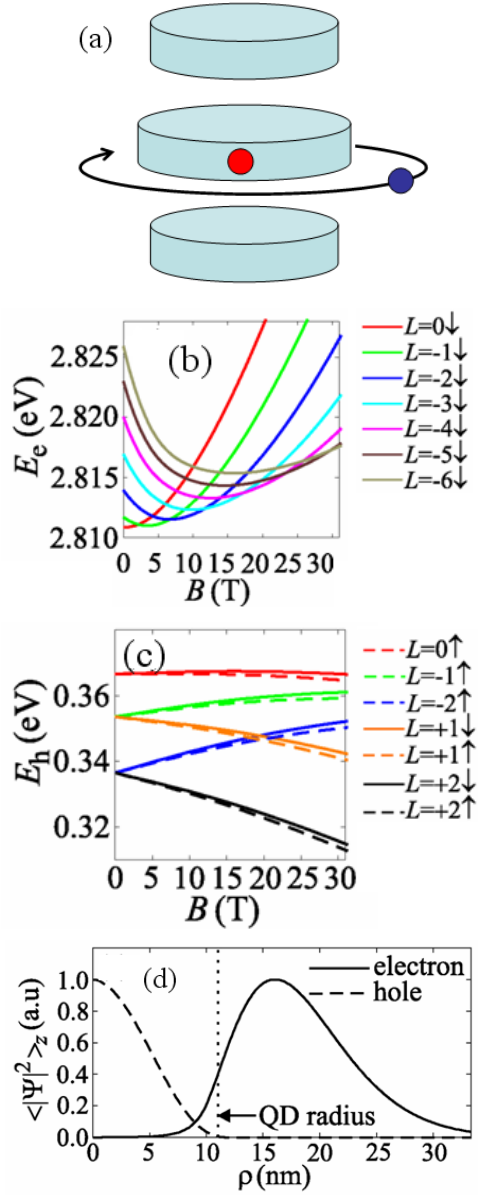

**FIG. 1.** (a) Model of stacked Type-II QDs with a hole localized within one dot and electrons free to move around the whole stack. (b) Dependence of the electron levels on magnetic field. (c) Dependence of the heavy hole levels on magnetic field. (d) The probability densities of the ground electron and hole states averaged over $z$, B = 0 T.

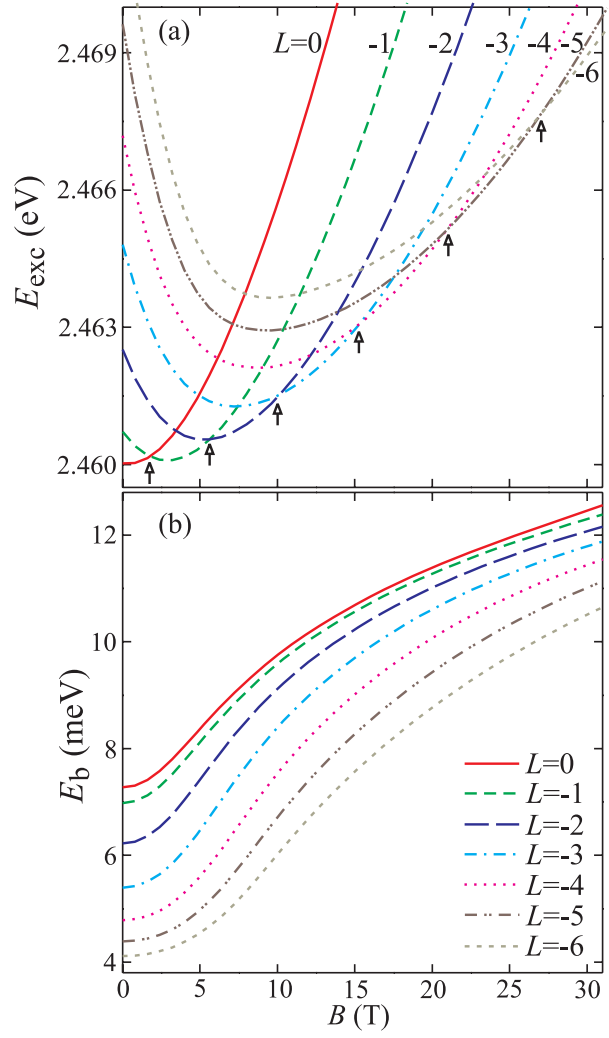

**FIG. 2.** (a) Dependence of the computed spin-down exciton energy levels on the magnetic field. (b) Binding energies of the lowest energy exciton states of different orbital momenta.

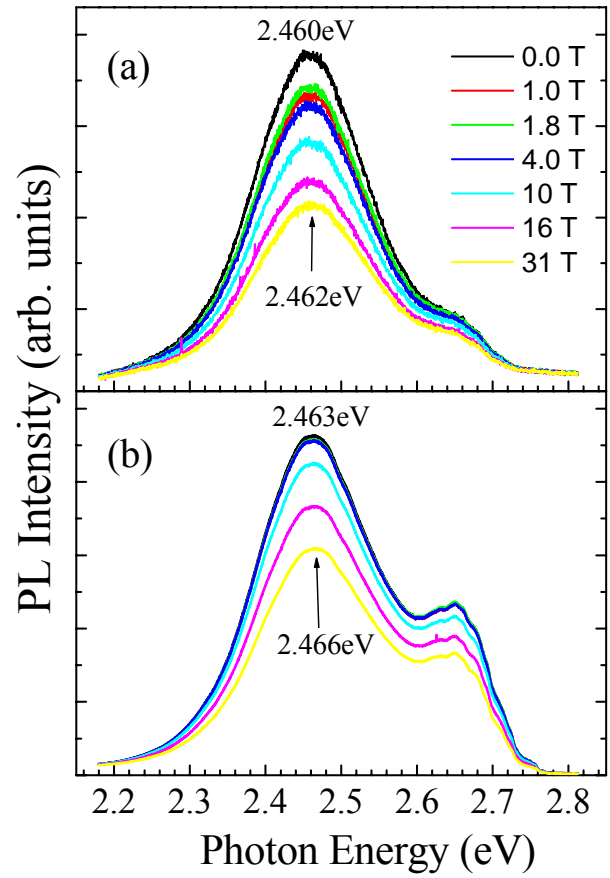

**FIG. 3.** Magneto-photoluminescence from ZnTe/ZnSe multiple type-II quantum dots: (a) the lowest excitation intensity, $I_0$; (b) the excitation intensity $100 \times I_0$.

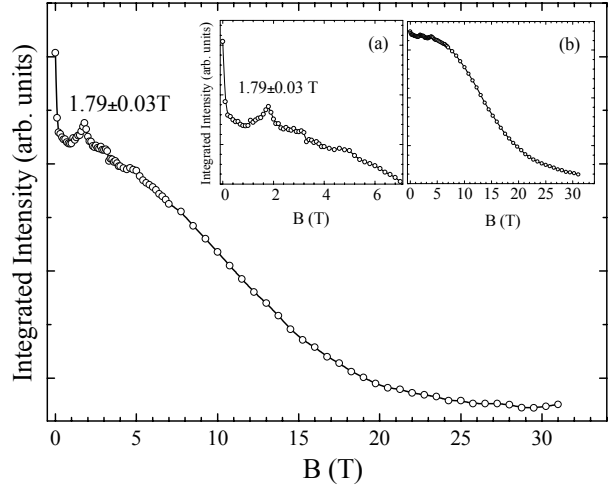

**FIG. 4.** Integrated magneto-photoluminescence from ZnTe/ZnSe multiple type-II quantum dots at the lowest excitation intensity, $I_0$. Inset (b) is the same for the excitation $100 \times I_0$. Inset (a) shows a low field region in greater details

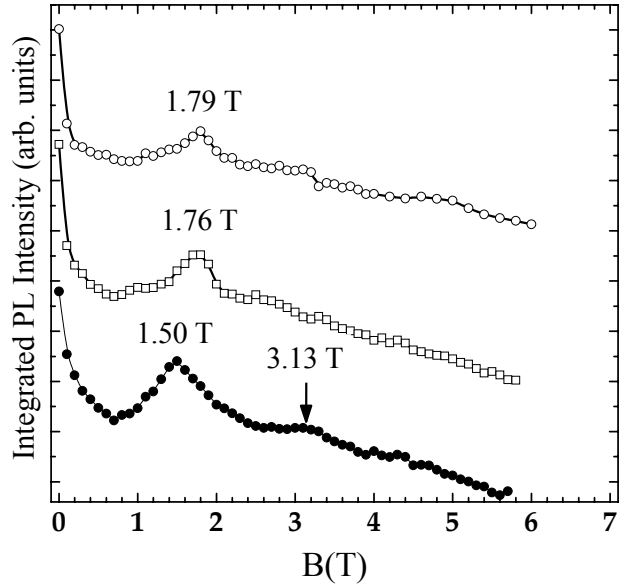

**FIG. 5.** Integrated magneto-photoluminescence from three ZnTe/ZnSe type-II quantum dot samples grown with increasing Te content (from top to bottom). The magnetic field of the angular momentum transition shifts to lower values. The peak at 3.1 T observed in the lower curve is due to elongation of quantum dots.